\documentclass[prl,superscriptaddress,twocolumn]{revtex4}

\usepackage{enumerate}
\usepackage{amsfonts,amssymb,amsmath}
\usepackage[]{graphics,graphicx,epsfig}
\usepackage{amsthm}

\usepackage{graphicx}
\usepackage{dcolumn}
\usepackage{natbib}
\usepackage{color}
\usepackage{multirow}
\usepackage{ulem}

\begin{document}

\title{Weighted Bures Length Uncovers Quantum State Sensitivity}

\author{Pawe{\l} Kurzy{\'n}ski}
\email{pawel.kurzynski@amu.edu.pl}
\affiliation{Institute of Spintronics and Quantum Information, Faculty of Physics, Adam Mickiewicz University, Uniwersytetu Pozna{\'n}skiego 2, 61-614 Pozna\'n, Poland}
\affiliation{Centre for Quantum Technologies,
National University of Singapore, 3 Science Drive 2, 117543 Singapore,
Singapore}

\date{\today}


\begin{abstract}

The unitarity of quantum evolutions implies that the overlap between two initial states does not change in time. This property is commonly believed to explain the lack of state sensitivity in quantum theory, a feature that is prevailing in classical chaotic systems. However, a distance between two points in classical phase space is a completely different mathematical concept than an overlap distance between two points in Hilbert space. There is a possibility that state sensitivity in quantum theory can be uncovered with a help of some other metric. Here we show that the recently introduced Weighted Bures Length (WBL) achieves this task. In particular, we numerically study a cellular automaton-like unitary evolution of $N$ qubits, known as Rule 54, and apply WBL to show that a single-qubit perturbation of a random initial state: (a) grows linearly in time under the nearest neighbour interaction on a cycle, (b) appears to grow exponentially in time under interaction given by a random bipartite graph. 

\end{abstract}

\maketitle




\section{Introduction}


An overlap between two quantum states $|\langle \psi_1|\psi_2 \rangle|$ is perhaps the most common measure of quantum state dissimilarity. Still, it fails to capture some intuitive differences in many-body systems. Imagine a collection of $N$ qubits and consider three different states
\begin{eqnarray}
|\psi_1\rangle &=& |000\ldots 0\rangle, \label{s1}\\
|\psi_2\rangle &=& |100\ldots 0\rangle, \label{s2}\\
|\psi_3\rangle &=& |111\ldots 1\rangle. \label{s3}
\end{eqnarray}  
It is quite clear that states $|\psi_1\rangle$ and $|\psi_2\rangle$ are much more alike than $|\psi_1\rangle$ and $|\psi_3\rangle$, or $|\psi_2\rangle$ and $|\psi_3\rangle$. Nevertheless,
\begin{equation}
|\langle \psi_1|\psi_2 \rangle| = |\langle \psi_1|\psi_3 \rangle| = |\langle \psi_2|\psi_3 \rangle| = 0.
\end{equation} 
The above simple example clearly motivates the search for other measures of quantum state dissimilarity capable of capturing such differences.

In addition, the overlap invariance under unitary evolutions 
\begin{equation}
|\langle \psi_1|U^{\dagger} U |\psi_2 \rangle| = |\langle \psi_1|\psi_2 \rangle|
\end{equation}
is commonly believed to be the reason why a quantum analogue of a classical state sensitivity to initial conditions is so elusive.  
This fact stimulated development of alternative approaches to quantum state sensitivity \cite{Haake}. For example, one can evolve the system forward in time, perturb it, and then evolve it backwards in time. In such a case the overlap between the initial state and the evolved forward -- perturbed -- evolved backward state does uncover some aspects of state sensitivity in continuous variable systems \cite{BZ}. A similar method, know as Loschmidt echo \cite{LE1,LE2,LE3,LE4}, can be used to study how unitary dynamics changes under small perturbations of the governing Hamiltonian.

Here we focus on a recently introduced measure of dissimilarity of multipartite quantum states  \cite{WBL} -- the  {\it Weighted Bures Length} (WBL). It is a metric $D_B(\rho,\sigma)$ that was particularly designed to deal with the problems exemplified by the states (\ref{s1}-\ref{s3}). For the above three states one gets
\begin{eqnarray}
D_B(|\psi_1\rangle,|\psi_2\rangle) &=& \frac{\pi}{2}, \\
D_B(|\psi_1\rangle,|\psi_3\rangle) &=& N \frac{\pi}{2}, \\
D_B(|\psi_2\rangle,|\psi_3\rangle) &=& (N-1) \frac{\pi}{2}, 
\end{eqnarray}
which exactly reflects our intuitions about these states. The goal of this work is to show that $D_B(\rho,\sigma)$ is also capable of detecting quantum state sensitivity.


\section{Weighted Bures Length}

Let us briefly discuss the main idea behind the derivation of the WBL \cite{WBL}. Consider an $N$-partite quantum system and two density matrices $\rho_N$ and $\sigma_N$, corresponding to two different states. The system is divided into parts according to a partition $P_{k_{\alpha}}$. The parts are labeled by $\alpha$ and $k_{\alpha}$ is the number of elements in a given part. For example, for $N=3$ the possible partitions can be schematically represented as
\begin{eqnarray}
& &\{1,2,3\}, \nonumber \\
& &\{1,\{2,3\}\},~\{2,\{1,3\}\},~\{3,\{1,2\}\}, \nonumber \\
& &\{\{1\},\{2\},\{3\}\}.
\end{eqnarray} 
The corresponding sizes of each partition are
\begin{eqnarray}
& &\{k_1 = 3\}, \nonumber \\
& &\{k_1 = 1,k_2=2\},~\{k_1 = 1,k_2=2\},~\{k_1 = 1,k_2=2\}, \nonumber \\
& &\{k_1 = 1,k_2=1,k_3=1\}.
\end{eqnarray} 
The reduced density matrices describing the state of each partition are $\rho_{k_{\alpha}}$ and $\sigma_{k_{\alpha}}$. The WBL between  $\rho_N$ and $\sigma_N$ is defined as 
\begin{equation}\label{WBL}
D_B(\rho_N,\sigma_N) = \max_{P_{k_{\alpha}}} \delta_{B,P_{k_{\alpha}}}(\rho_N,\sigma_N),
\end{equation}
where 
\begin{equation}
\delta_{B,P_{k_{\alpha}}}(\rho_N,\sigma_N) = \sum_{\alpha} \frac{1}{k_{\alpha}} B(\rho_{k_{\alpha}},\sigma_{k_{\alpha}}),
\end{equation}
and
\begin{equation}
B(\rho_{k_{\alpha}},\sigma_{k_{\alpha}}) = \cos^{-1}F(\rho_{k_{\alpha}},\sigma_{k_{\alpha}})
\end{equation}
is the standard Bures distance \cite{Bures} based on the quantum fidelity \cite{F1,F2,F3}
\begin{equation}
F(\rho_{k_{\alpha}},\sigma_{k_{\alpha}}) = \text{Tr} \sqrt{ \sigma_{k_{\alpha}}^{1/2} \rho_{k_{\alpha}}\sigma_{k_{\alpha}}^{1/2} }.
\end{equation}

The evaluation of WBL (\ref{WBL}) requires optimisation over all possible partitions. The exact values for some specific classes of $N$-qubit states were presented in \cite{WBL}. For the purpose of this work we list one of them
\begin{eqnarray}
D_B\left( |0\rangle^{\otimes N}, |ghz_k\rangle \otimes |0\rangle^{\otimes (N-k)}  \right) = k \cos^{-1}|a|, \label{d2}
\end{eqnarray}
where
\begin{equation}\label{ghz}
|ghz_k\rangle = a |0\rangle^{\otimes k} + b |1\rangle^{\otimes k}
\end{equation}
is the Greenberger-Horne-Zeilinger (GHZ) state \cite{GHZ}.

Apart from being a metric, an important feature of WBL is its contractivity. WBL does not increase under completely positive trace-preserving operations performed on a single subsystem \cite{WBL,PC}. However, as we are going to show, WBL increases under particular multi-partite operations.


\section{Rule 54}

We focus on a N-qubit cellular automaton-like unitary dynamics known as {\it Rule 54} \cite{R54c1}. Its name originates from the Wolfram code \cite{Wolfram} that assigns a unique number to every one-dimensional two-state cellular automaton. Each qubit has two neighbours, two other qubits with which it interacts, and the dynamics flips the state of the qubit if at least one of its two neighbours is in the state $|1\rangle$. The corresponding three-qubit transformation is given by
\begin{eqnarray}
U_x &=& |101\rangle\langle 111| + |110\rangle\langle 110| + |111\rangle\langle 101| \nonumber \\
&+& |110\rangle\langle 100| + |001\rangle\langle 011| + |010\rangle\langle 010| \nonumber \\
&+& |011\rangle\langle 001| + |000\rangle\langle 000|. \label{rule}
\end{eqnarray}
In the above $|n_1~ x~ n_2 \rangle$ describes the state of the target qubit ($x$) and the states of its neighbours ($n_1$ and $n_2$).

Rule 54 is a perfect testbed for many-body dynamics. It has been successfully applied to study various aspects of multipartite systems, such as nonequilibrium steady states \cite{R54Integ}, thermalisation \cite{R54ETH1,R54ETH2} or operator entanglement spreading \cite{R54Ent1,R54Ent2}.  It was proven to be integrable when the dynamics is defined on a chain \cite{R54c1}, i.e., the qubits are labeled by integers ($x \in {\mathbb Z}$) and a single step of dynamics is given by
\begin{eqnarray}
U = \left(\prod_{x~odd} U_x \right) \times \left(\prod_{x~even} U_x \right), \label{chain}
\end{eqnarray}
where the neighbours of qubit $x$ are labeled $x\pm 1$. 

Here we apply Rule 54 to a collection of $N$ qubits whose interactions are determined by a bipartite graph. In a bipartite graph the vertices are divided into two disjoint sets, $\mathbb{V}_1$ and $\mathbb{V}_2$, and the set of edges/arcs $\mathbb{E}$ consists of elements $(v_1,v_2)$ and $(u_2,u_1)$ such that $v_1,u_1 \in \mathbb{V}_1$ and $v_2,u_2 \in \mathbb{V}_2$. Note, that a chain graph is a special case of a bipartite graph in which $\mathbb{V}_1$ is the set of vertices with odd labels and $\mathbb{V}_2$ is the set of vertices with even labels. 

\begin{figure}
	\includegraphics[width=8.5cm]{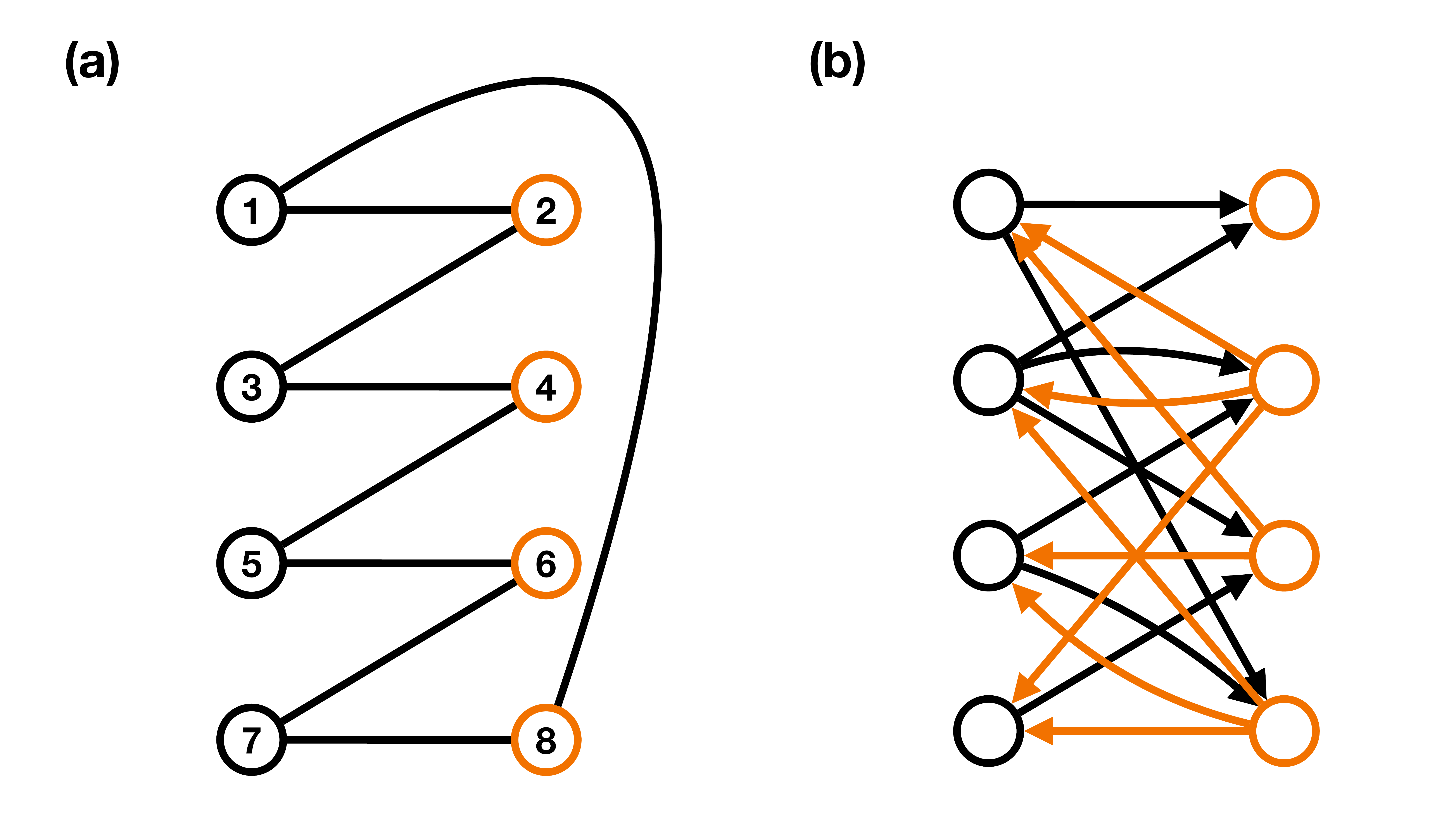}
  \caption{Schematic representation of the two scenarios considered in this work: (a) Rule 54 applied to N-cycle graph; (b) Rule 54 applied to a random bipartite graph.} \label{f1}
\end{figure}

The above means that in our model the qubits are divided into two sets. If qubit $x$ is in $\mathbb{V}_1$ ($\mathbb{V}_2$), its neighbours are in $\mathbb{V}_2$ ($\mathbb{V}_1$). Therefore, in our case a single step of the evolution is given by
\begin{eqnarray}
U = \left(\prod_{v\in \mathbb{V}_2} U_v \right) \times \left(\prod_{v \in \mathbb{V}_1} U_v \right), \label{evol}
\end{eqnarray}
We consider two particular scenarios: (a) the qubits are arranged into N-cycle and the interaction occurs between the nearest neighbours just like in Eq. (\ref{chain}); (b) the interaction between qubits is described by a random bipartite graph, i.e., each qubit randomly chooses two neighbours from the opposite set. Note, that each vertex has exactly two incoming arcs (exactly two neighbours), but the number of outgoing arcs may differ from two (each qubit can be a neighbour to more than two, or less than two, qubits from the opposite set). These two scenarios are schematically represented in Fig. \ref{f1}.

\section{Results}

The initial state of the system $|\psi (0)\rangle$ is a random basis state, such as $ |001010\ldots \rangle$. In simple words, it corresponds to a random classical bit string of length $N$. The perturbation of this state is chosen to be a single-qubit unitary transformation
\begin{eqnarray}
V|0\rangle &=& a|0\rangle + b|1\rangle, \label{pert1} \\
V|1\rangle &=& a|1\rangle - b|0\rangle, \label{pert2}
\end{eqnarray}
applied to a randomly chosen qubit. After the perturbation the state becomes $|\tilde{\psi} (0)\rangle$. 
Next, we numerically study the evolution of both states in scenarios (a) and (b) and analyse 
\begin{equation}
{\mathcal{D}}(t) \equiv D_B(|\psi (t)\rangle,|\tilde{\psi} (t)\rangle). 
\end{equation}

To evaluate ${\mathcal{D}}(t)$ we make the following observation. Rule 54 transforms basis states into basis states -- it does not generate superpositions. Therefore, the state $|\psi (t)\rangle$ is a basis state for all $t$. On the other hand, the state  $|\tilde{\psi} (t)\rangle$ is a superposition of two basis states \begin{equation}
|\tilde{\psi} (t)\rangle = a |\psi (t)\rangle + b |\psi_{\perp} (t)\rangle, 
\end{equation}
where $|\psi_{\perp} (t)\rangle$ is a basis state orthogonal to $|\psi (t)\rangle$. Next, note that WBL does not change under single-qubit operations \cite{WBL,PC}, hence one can always transform the local bases such that
\begin{eqnarray}
|\psi(t)\rangle &\rightarrow& |0\rangle^{\otimes N},  \label{t1} \\
|\tilde{\psi} (t)\rangle &\rightarrow& |ghz_{k_t}\rangle \otimes |0\rangle^{\otimes(N-k_t)}, \label{t2}
\end{eqnarray}
where $|ghz_{k_t}\rangle$ is given by (\ref{ghz}) and $k_t$ is the time-dependent number of positions at which $|\psi_{\perp} (t)\rangle$ differs from $|\psi (t)\rangle$. In fact, $k_t$ is the Hamming distance between the bit strings corresponding to $|\psi_{\perp} (t)\rangle$ and $|\psi (t)\rangle$. As a result, Eq. (\ref{d2}) implies that
\begin{eqnarray}
{\mathcal{D}}(t) &=& D_B\left( |0\rangle^{\otimes N},|ghz_{k_t}\rangle \otimes |0\rangle^{\otimes(N-k_t)}\right) \nonumber \\ 
&=& k_t \cos^{-1} |a|. 
\end{eqnarray}
Since $k_0 = 1$ and $k_t \geq k_0$, we have that ${\mathcal{D}}(t) \geq {\mathcal{D}}(0)$. We see that it is enough to focus on $k_t$ and the goal is to estimate its growth rate.

\begin{figure}
	\includegraphics[width=8.5cm]{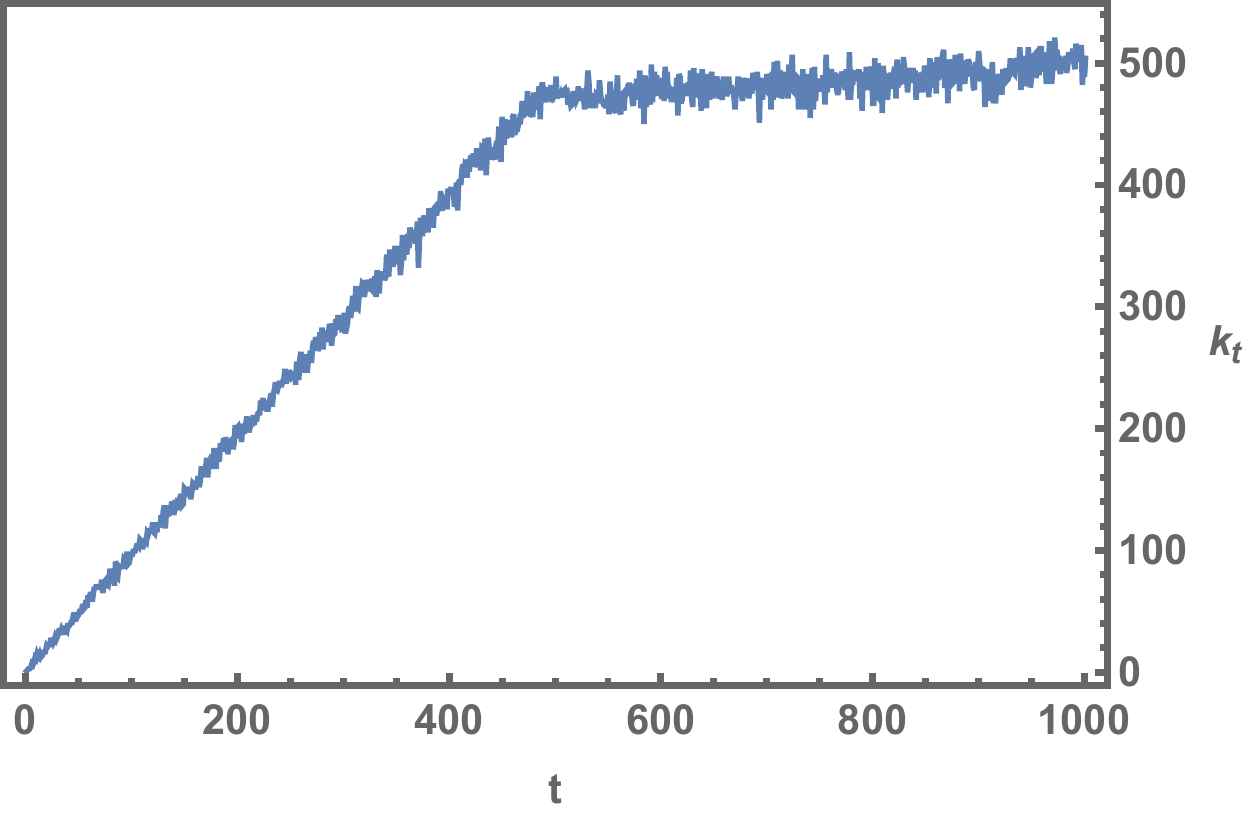}
  \caption{An example of the evolution of $k_t$ for scenario (a). The number of qubits is $N=1000$. The points were connected for a better visualisation.} \label{f2}
\end{figure} 

In case of scenario (a) a single-qubit perturbation spreads to nearest neighbours, therefore the growth of $k_t$ can be at most linear in $t$. Indeed, this is confirmed by numerical simulations (see Fig. \ref{f2}). The value of $k_t$ grows linearly in time till it reaches approximately $N/2$. The finite growth is caused by the finite size of the system. The value of $N/2$ is expected since it corresponds to the average Hamming distance between two random bit strings of length $N$. After some time the value of $k_t$ starts to drop. This is again due to the finite size of the system and due to unitarity of the dynamics. More precisely, Rule 54 is a permutation, therefore the system must return to the initial state after a finite number of steps.  

\begin{figure}
	\includegraphics[width=8.5cm]{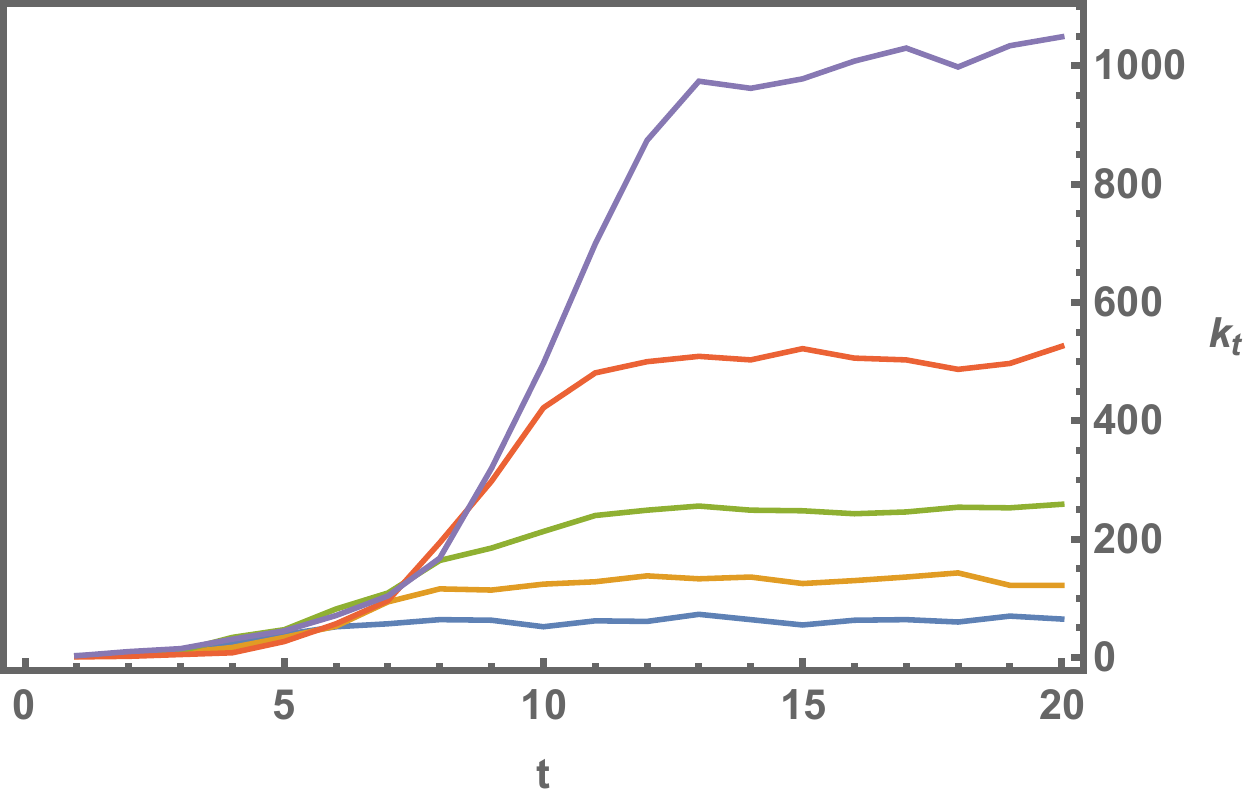}
  \caption{An example of the evolution of $k_t$ for scenario (b). The number of qubits is: $N=128$ (blue), $N=256$ (orange), $N=512$ (green), $N=1024$ (red), $N=2048$ (purple). The points were connected for a better visualisation.} \label{f3}
\end{figure} 

The growth of $k_t$ can be much faster than linear if the neighbourhood is chosen randomly -- scenario (b). Numerical simulations show that the value of $N/2$ is reached just after few steps (see Fig. \ref{f3}). We observed for few values of $N$ ($N=128,254,512,1024,2048$) that on the average after $\log N$ steps $k_t \geq N/4$, which suggests an exponential growth of $k_t$ during the first stage of the evolution. However, a more detailed analysis of this scenario is needed to confirm this conjecture. It looks like scenario (b) exhibits exponential sensitivity to initial conditions, which in classical terms would be interpreted as a chaotic behaviour.  


\section{Conclusions}

We showed that WBL \cite{WBL} can be used to study quantum state sensitivity. In particular, we performed numerical studies of a $N$-qubit dynamics governed by Rule 54 \cite{R54c1} and showed that a single qubit perturbation, given by Eqs. (\ref{pert1}) and (\ref{pert2}), grows in time. More precisely, we observed that if the interaction between the qubits is governed by N-cycle graph, the WBL between the initial state and the perturbed state grows linearly in time. On the other hand, we found that, if the interaction between the qubits is governed by a random bipartite graph, the WBL between the initial state and the perturbed state appears to grow exponentially in time. 

There are a few avenues of research stemming from this work. It would be interesting to apply WBL to study quantum state sensitivity in other many-body systems. In addition, it is natural to look for the relation between the quantum state sensitivity uncovered by WBL and commonly accepted measures of quantum chaotic behaviour, such as Loschmidt echo \cite{LE1,LE2,LE3,LE4} or out-of-time-order correlator (OTOC) \cite{OTOC1,OTOC2,OTOC3,OTOC4}.


\section{Acknowledgements}

This research is supported by the Polish National Science Centre (NCN) under the Maestro Grant no. DEC-2019/34/A/ST2/00081. 



\end{document}